\documentclass{jpconf}

\usepackage{graphicx}

\begin{document}

\title{Electron spin relaxation in organic semiconductors probed through $\mu$SR}

\author{L Nuccio $^1$, L Schulz $^2$, M Willis $^1$, F L Pratt $^3$, M Heeney $^4$, N Stingelin $^4$, C Bernhard $^2$, A J Drew  $^{1,2}$}

\address{$^1$ Queen Mary University of London, Department of Physics, Mile End Road, London, E1 4NS, UK}
\address{$^2$ Department of Physics \& FriMat, University of Fribourg, Ch. du Mus\'{e}e 3, 1700 Fribourg, CH}
\address{$^3$ ISIS Muon Facility, Rutherford Appleton Laboratory, Didcot, OX11 0QX, UK}
\address{$^4$ Centre for Plastic Electronics, Imperial College London, Exhibition Road, London, SW7 2AZ, London, UK}

\ead{l.nuccio@qmul.ac.uk}

\begin{abstract}
Muon spin spectroscopy and in particular the avoided level crossing technique is introduced, with the aim of showing it as a very sensitive local probe for electron spin relaxation in organic semiconductors. Avoided level crossing data on TMS-pentacene at different temperatures are presented, and they are analysed to extract the electron spin relaxation rate, that is shown to increase on increasing the temperature from 0.02 MHz to 0.33 MHz at 3 K and 300 K respectively.
\end{abstract}



\section{Introduction}
Organic semiconductors are extremely promising materials for spintronics thanks to their long spin coherence times. However, spin phenomena in these materials are not yet fully understood, partly due to the lack of suitable characterisation techniques, since those tuned for inorganic semiconductors are not always successful for organic materials \cite{sanvito_2007,drew_2009}. Muon spin relaxation ($\mu$SR) has been proven to be an applicable probe in this field, as it is sensitive to spin dynamics in organic semiconductors on a molecular lengthscale \cite{drew_2008}. In organic semiconductors the implanted muons form a hydrogen-like system, called muonium, by capturing an electron. In muonium the muon  and electron spins are coupled through hyperfine interaction. As a consequence relaxation on the electron spin can induce changes in the muon spin, which can be detected. Here we show how muon spin relaxation and in particular the avoided level crossing (ALC) technique can be used to probe the electron spin dynamics in organic semiconductors.

\section{Muon spin relaxation and avoided level crossings}
\begin{figure}[htbp]
\begin{center}
\includegraphics[width=140mm]{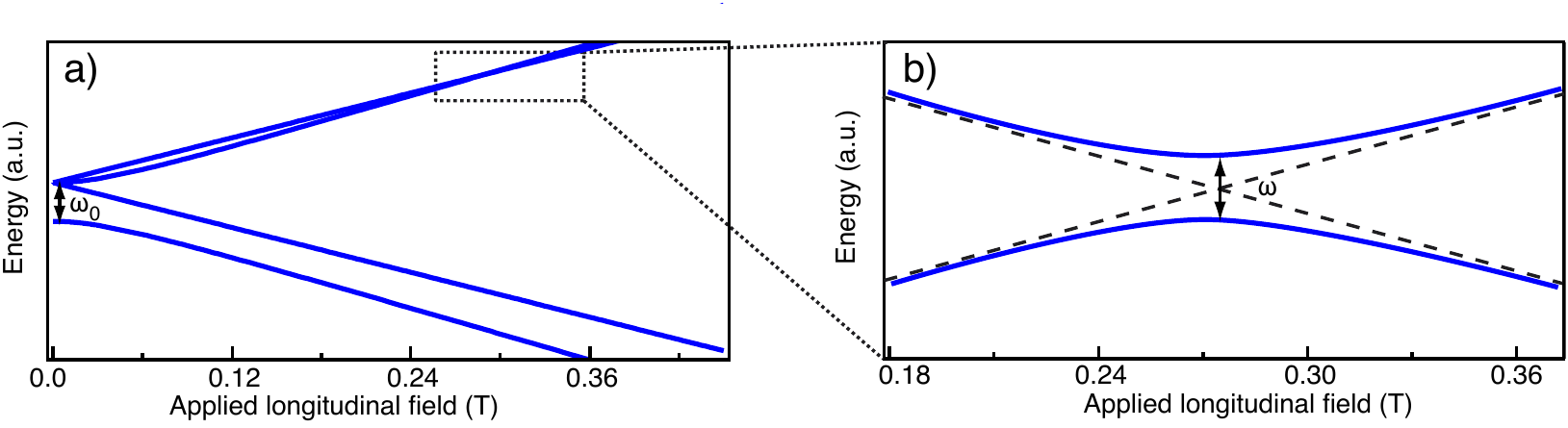}
\caption{(a) The muonium Breit-Rabi diagram with a hyperfine coupling constant of A=80 MHz (b) Zoom-in region of the energy levels near the ALC.}
\label{fig1}
\end{center}
\end{figure}

\begin{figure}[htbp]
\begin{center}
\includegraphics[width=90mm]{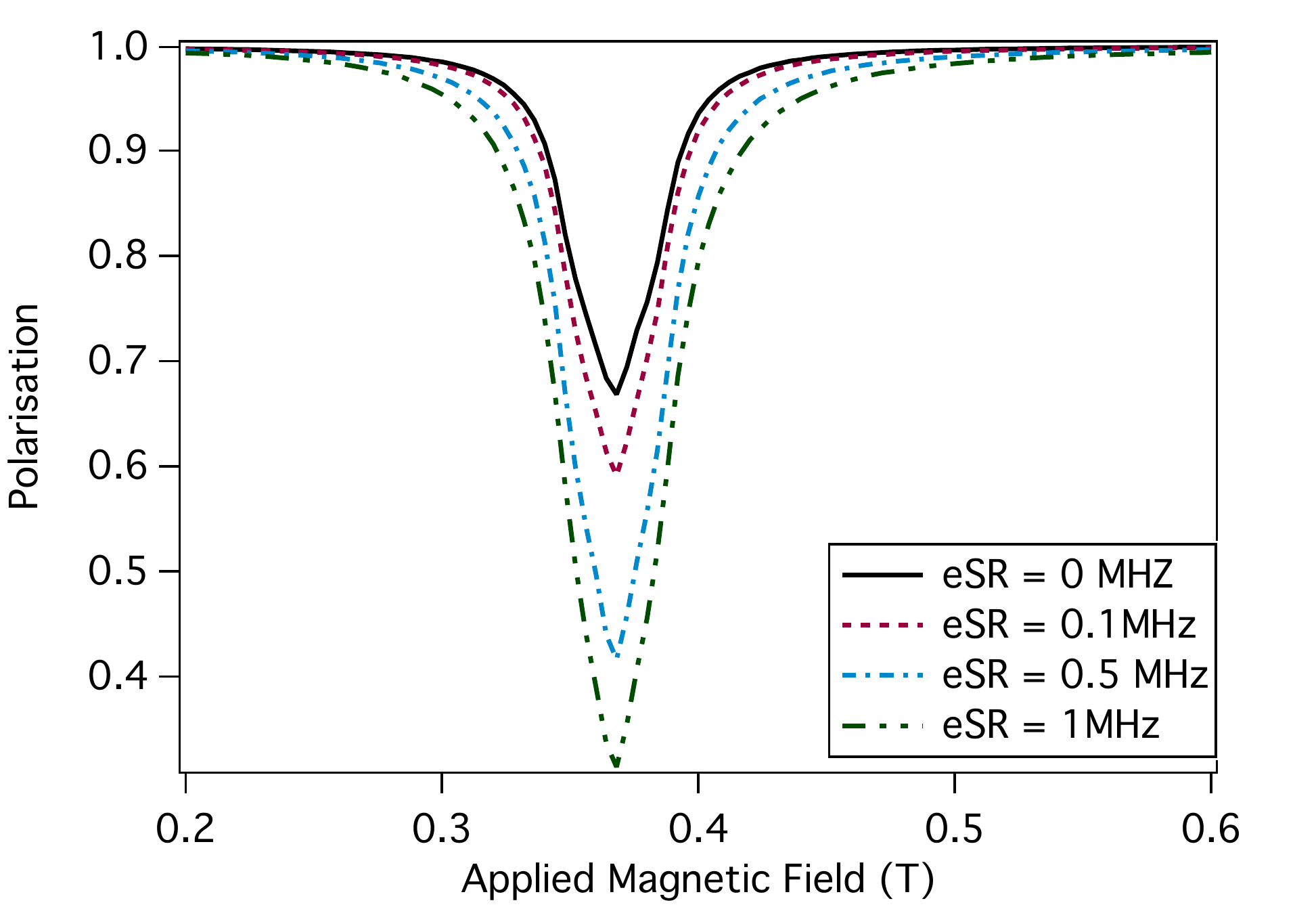}
\caption{Simulated ALC resonance for different electron spin relaxation rates (eSR)}
\label{fig2}
\end{center}
\end{figure}

Muon spin relaxation is an experimental technique using positive muons as microscopic spin probe. Positive muon has a finite lifetime $\tau\sim$2.2 $\mu$s, and it radioactively decays emitting a positron and two neutrinos according to the process $\mu^+\rightarrow e^++\nu_e+\overline{\nu_{\mu}}$. The positron is emitted preferentially in the direction of the muon's spin at the time of the decay, due to parity violation in weak interaction \cite{muonbook,blundell_2004}.  As a consequence measuring the time dependence of the emission direction of the positron provides information on the time evolution of the muon spin.
\\
In $\mu$SR experiments 100\% spin polarised muons are implanted in the sample. Once implanted in semiconductors the muon gives rise to a series of ionisation and scattering processes, through which its energy decreases from 4MeV to a few keV. Then it takes part in a series of electron capture and loss processes that reduce its energy further. Finally it thermalises to form interstitial positive muon or it captures an electron and form a hydrogen-like species called muonium (Mu) \cite{blundell_2004}. Thermalisation of muons is fast ($<$ns) compared to its lifetime.
\\
If an external magnetic field is applied in the direction of the initial muon spin, as it is in so called “longitudinal field” experiments, the Hamiltonian for muonium can be written as
\begin{equation}
\label{hamiltonian}
H=-\gamma_\mu\hbar\bi{I}\cdot\bi{B}+\gamma_e\hbar\bi{S}\cdot\bi{B}+\hbar\bi{S}\cdot\bi{A}\cdot\bi{I}
\end{equation}
where $\gamma_\mu$ and $\gamma_e$ are the muon and electron gyromagnetic ratios, B is the applied longitudinal field, I and S are the muon’s and electron’s spins and A is the hyperfine tensor \cite{blundell_2004}. 
The energy levels as a function of the applied magnetic field that originates from this Hamiltonian for an isotropic hyperfine interaction are shown in the Breit-Rabi digram in Figure \ref{fig1}(a).
 At B=0 a singlet and a triplet are obtained, the degeneracy of the latter being lifted by the magnetic field. At high enough fields (Paschen-Back regime, where most of the experiments are performed) the electron and muon spins are decoupled and the state of the system can be described by products of pure Zeeman states for the electron and muon. 
\\
At a high field which depends on the value of the hyperfine coupling constant, two of the levels should cross, as can be observed in Figure  \ref{fig1}(b) (dashed lines). 
However, if further interactions are present, as an example with the nuclei, this crossing can be avoided. In fact, for a Hamiltonian $H=H_0+H^I$ where $H^I$ can be treated as a perturbation, the correction to the energies up to second order is given by
\begin{equation}
\label{perturbation}
E_n-E_n^0=\langle\chi_n|H^I|\chi_n\rangle+\sum_m\frac{|\langle\chi_n|H^I|\chi_n\rangle|^2}{E_n^0-E_m^0}
\end{equation}
where $E_n^0$ and $\chi_n$ are the n-th zero order energy and eigenstate and the denominator becomes tiny when the two unperturbed levels approach \cite{muonbook}. The effect of this correction to the energy levels is to avoid the crossing (See Figure  \ref{fig1}(b)). Close to this avoided level crossing the states are no longer pure Zeeman states, but they are instead mixtures of states corresponding to different values of the muon magnetic quantum number. As a consequence a loss of polarisation of the muon can be observed \cite{muonbook,blundell_2004}. ALCs can be classified by the change in total quantum number M. If the hyperfine interaction is isotropic, the most common kind of ALC is the one with $\Delta$M=0, that corresponds to a muon-nuclear spin flip-flop. However in solids the most intense crossing is $\Delta$M=1, corresponding to a muon spin flip and is observed only when anisotropy in the interaction is present. $\Delta$M=2 transitions exist, but they are usually very weak and narrow [3].
A simulated ALC curve is shown in Figure \ref{fig2} (black line). 
\newline
As can be observed, the polarisation is plotted versus the applied magnetic field. The polarisation exhibits a deepening, representing the loss of polarisation around the ALC field.  The shape and position of this ALC curve depends on the hyperfine coupling constants (both isotropic and anisotropic). However, the presence of an electron spin relaxation rate (eSR) also has an effect on the ALC curve. ALCs simulated using the same parameters with an increasing electron spin relaxation rate are shown in Figure  \ref{fig2}. It is clear that at low electron spin relaxation rates neither the position nor the shape of these ALCs show substantial changes, but the amplitude enormously increases. 
\\
This can be modeled through a density matrix formalism, that is also used to simulate the experimental data and extract the eSR. A density matrix is defined as
\begin{equation}
\label{densmat}
\rho(t)=\frac{1}{4}[1+\bi{p}(t)\cdot\bi{\sigma}+\bi{p}_e(t)\cdot\bi{\tau}+\sum_{j,k}P^{jk}(t)\sigma^j\tau^k]
\end{equation}
where $\bi{p}(t)$, $\bi{p}_e(t)$  and $P^{jk}(t)$ are the muon, electron and mixed polarisations, respectively. 
The time evolution of $\rho(t)$ is described by the equation
\begin{equation}
\label{timeevrho}
i\hbar\frac{d\rho}{dt}=[H,\rho]
\end{equation}
from which the time dependent polarisations can be obtained \cite{patterson_1988}. The general solution can only be determined numerically. If an electron spin relaxation mechanism is present in the system its effect can be taken into account phenomenologically by introducing extra terms $-\lambda p_e$  and $-\lambda p^{jk}$  in the differential equations for the polarisations obtained from equation \ref{timeevrho} \cite{meier_1982}:

\begin{equation}
\label{fenrel1}
\frac{dp_e}{dt}=...-\lambda p_e     ,      \frac{dP^{jk}}{dt}=...-\lambda P^{jk}
\end{equation}


\section{Experimental methods}
\begin{figure}[htbp]
\begin{center}
\includegraphics[width=40mm]{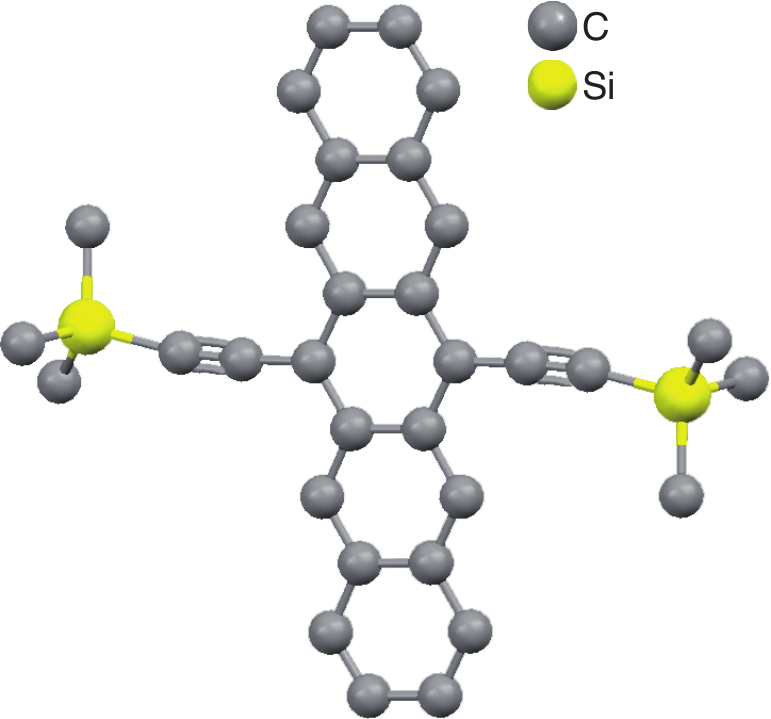}
\caption{Molecular structure of TMS-pentacene. Hydrogen has been omitted for clarity.}
\label{fig3}
\end{center}
\end{figure}

(Trimethyl)silyl-pentacene (TMS-pentacene) was synthesised according to a published procedure \cite{anthony_2002,anthony_2006} and was further purified by repeated recrystallisation from dichloromethano/ethanol. Its structure is shown in Figure \ref{fig3}. 
The presence of the side groups on the central spine prevents the molecule to take the usual herringbone packing of pentacene; as a consequence the molecules have a face-to-face packing that allows better $\pi$-orbitals overlap and consequently better charge carrier mobility \cite{anthony_2002,anthony_2006}.  The measured sample was polycrystalline powder.
\\
Muon experiments were performed at the HIFI spectrometer at ISIS, Rutherford Appleton Laboratory. About 250 mg of TMS-pentacene was placed in an envelope (17mm $\times$ 17mm) made of 25 $\mu$m thick silver foil (99.99 \% pure). Another layer of silver foil was placed in front of the sample as a degrader, to assure that the muons stop in the sample.
\\
The software Wimda \cite{pratt_2000} was used to extract the time-integrated data and plot them as a function of applied field. ALCs were obtained in this way, and they were then modelled using the software Quantum based on the density matrix formalism described in the previous section \cite{lord_2006}. The ALC simulations presented in the following are performed considering a muon-electron system. No nuclei are taken into account as their influence was discovered to be negligible. Monte-Carlo simulations with powder averaging and 10000 iterations for each magnetic field was used.

\section{Results and discussion}
\begin{figure}[htbp]
\begin{center}
\includegraphics[width=90mm]{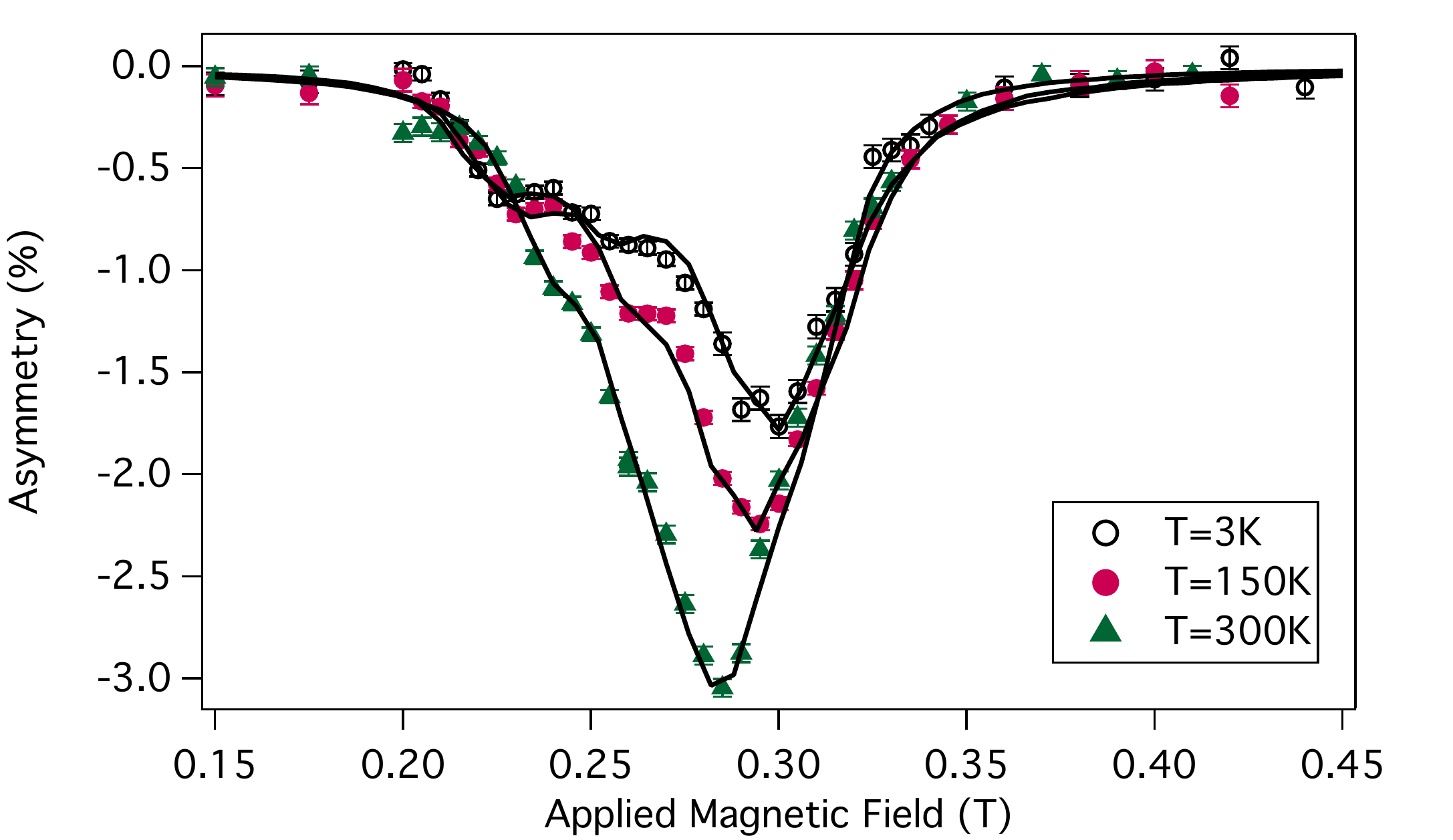}
\caption{Muon data around the ALC. Black lines are the result of modelling.}
\label{fig4}
\end{center}
\end{figure}

Figure \ref{fig4} shows the ALCs measured for TMS-pentacene at 3K, 150K and 300K. Data were also acquired at 10K and 25K (data not shown). ALCs are observed around 0.28T, corresponding to a hyperfine coupling of around 70MHz. The position and width of the ALCs do not show a remarcable temperature depence, while the amplitude considerably increases on increasing the temperature.  From the time-differential data and the field dependent relaxation rate (data not shown, see ref \cite{schulz_2010}) this increased ALC depth is clearly related to a higher relaxation rate of the muon spin.

\begin{figure}[htbp]
\begin{center}
\includegraphics[width=80mm]{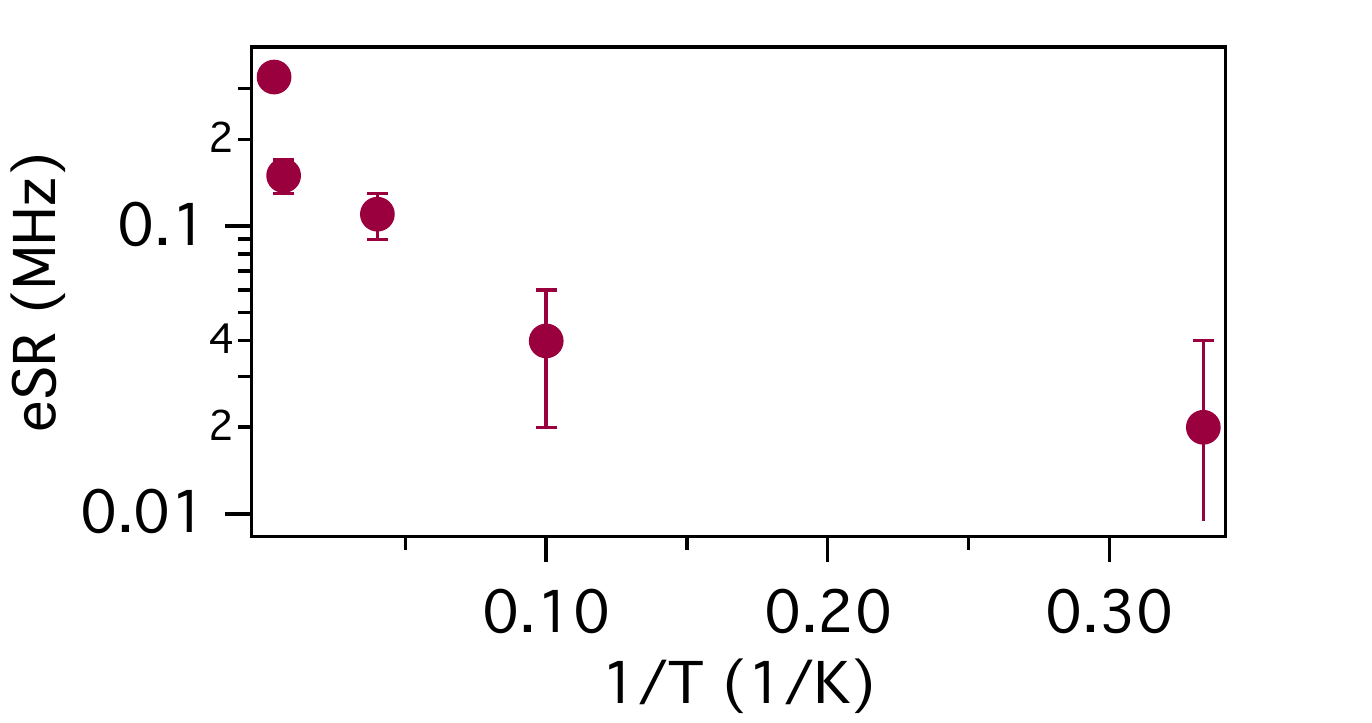}
\caption{Temperature dependence of the electron spin relaxation rate obtained from the simulations of the ALCs.}
\label{fig5}
\end{center}
\end{figure}

The measured ALCs look as the superposition of three ALCs very close to each other, the difference arising from slightly different possible bonding angles of the muon in the molecule. As a consequence three contributions were also considered in the modeling. Their relative amplitude is determined from the 3K data, setting the eSR to a very low value. This assumption is justified if the electron spin relaxation is driven by an effect strongly dependent on temperature, as a coupling to vibrations \cite{schulz_2010}. The result is shown in Figure \ref{fig4} (black line).The obtained relative amplitudes are 0.17, 0.24 and 0.59. These parameters are used to scale the three contributions at higher temperatures. An eSR different from zero is necessary in the model to account for the higher intensity of the ALC at higher temperatures, whereas there is only a small change in the hyperfine coupling constants. The results of these simulations are shown as solid lines in Figure \ref{fig4}. Values for hyperfine coupling constants A, D1 and D2 \cite{lord_2006} and for eSR as obtained from these simulations are reported in Table \ref{tabella}
\begin{table}
\caption{\label{tabella}Simulation parameters for TMS-pentacene.}
\begin{tabular}{@{}lllllll}
\br
 &\textbf{3K}& & &\textbf{10K}& & \\
\mr
 &ALC1&ALC2&ALC3&ALC1&ALC2&ALC3\\
A (MHz)&61.7$\pm$0.2&71.2$\pm$0.2&81.9$\pm$0.2&61.7$\pm$0.2&71.2$\pm$0.2&81.9$\pm$0.2\\
D1 (MHz)&3.6$\pm$0.3&6.0$\pm$0.4&5.8$\pm$0.4&3.6$\pm$0.3&6$\pm$0.4&5.0$\pm$0.4\\
D2 (MHz)&3.6$\pm$0.5&3.0$\pm$0.4&4.0$\pm$0.4&3.6$\pm$0.5&3.0$\pm$0.5&4.0$\pm$0.5\\
eSR (MHz)&0.02$\pm$0.02&0.02$\pm$0.02&0.02$\pm$0.02&0.04$\pm$0.02&0.04$\pm$0.02&0.04$\pm$0.02\\
\mr
&\textbf{25K}& & &\textbf{150K}& & \\
\mr
 &ALC1&ALC2&ALC3&ALC1&ALC2&ALC3\\
A (MHz)&62.5$\pm$0.2&71.8$\pm$0.2&81.6$\pm$0.2&63.0$\pm$0.2&72.2$\pm$0.2&81.0$\pm$0.2\\
D1 (MHz)&3.6$\pm$0.3&4.5$\pm$0.4&5.0$\pm$0.4&3.6$\pm$0.3&4.5$\pm$0.5&6.0$\pm$0.4\\
D2 (MHz)&3.6$\pm$0.5&2.0$\pm$0.5&4.0$\pm$0.5&3.6$\pm$0.5&2.0$\pm$0.5&4.0$\pm$0.5\\
eSR (MHz)&0.11$\pm$0.02&0.11$\pm$0.02&0.11$\pm$0.02&0.13$\pm$0.02&0.13$\pm$0.02&0.15$\pm$0.02\\
\mr
&\textbf{300K}& & & & & \\
\mr
 &ALC1&ALC2&ALC3& & & \\
A(MHz)&66.0$\pm$0.2&73.0$\pm$0.2&79.0$\pm$0.2& & & \\
D1(MHz)&3.6$\pm$0.3&4.5$\pm$0.4&6.0$\pm$0.4& & & \\
D2(MHz)&3.6$\pm$0.5&2.0$\pm$0.5&4.0$\pm$0.5& & & \\
eSR(MHz)&0.33$\pm$0.02&0.33$\pm$0.02&0.33$\pm$0.02& & & \\
\br
\br
\end{tabular}
\end{table}
The three ALCs tend to merge on increasing the temperature. These small changes in the values of A, D1 and D2, of the order of few percent, are likely to be due to changes in bond length at different temperatures. On the other hand there is a substantial increase in the eSR as a function of temperature, as can be seen in Table \ref{tabella} and in Figure \ref{fig5}. Figure \ref{fig5} shows the electron spin relaxation obtained from the simulation as a function of inverse temperature. This behaviour closely resembles the one observed in other organic semiconductors, like rubrene, Gaq$_{3}$ and TIPS-pentacene \cite{schulz_2010}. The electron spin relaxation rate in rubrene and TIPS-pentacene exhibits an Arrhenius-like temperature dependence, with two characteristic energy scales comparable to the energy of the vibrations in these systems  \cite{schulz_2010}. Data presented here are not detailed enough to extract a full temperature dependence of the eSR, but they are very consistent with \cite{schulz_2010}. It is also clear that the ALC spectroscopy is a very sensitive technique for the study of the electron spin relaxation in organic semiconductors. Further experiments are needed to get a detailed temperature dependence of eSR in TMS-pentacene. 

\section{Conclusions}
The present paper reviews the basic physics behind the avoided level crossing technique. It was shown that muons are a very effective probe to measure electron spin relaxation in organic semiconductors thanks to their coupling to the electron via hyperfine interaction. The electron spin relaxation rate can in fact be extracted from ALC measurements. As an example data on TMS-pentacene at three different temperatures are shown and analysed to obtain the electron spin relaxation rate, that is shown to increase on increasing the temperature.

\section*{Acknowledgements}
We would like to thank E. Roduner, Z. Salman and R. Scheuermann for valued discussions with regards to the analysis and interpretation of our data. AJD acknowledges financial support from the EPSRC (grant EP/G054568/1) and the Leverhulme Trust. CB acknowledges financial support from the SNF (grant 200020-119784 and 200020-129484) and the NCCR program MaNEP.

\end{document}